\documentclass{pasj00}

\def\commenta{$^*$}
\def\commentb{$^\dagger$}
\def\commentc{$^\ddagger$}
\def\commentd{$^\S$}

\def\inpress{in press}

\def\arxiv#1{ (arXiv astro-ph/#1)}

\DeclareAbbreviation\AAHam{Astron. Abh. Hamburg. Sternw.}
\DeclareAbbreviation\AARv{Astron. Astrophys. Rev.}
\DeclareAbbreviation\AAS{American Astron. Soc. Meeting Abstracts}
\DeclareAbbreviation\AcA{Acta Astron.}
\DeclareAbbreviation\actaa{Acta Astron.}
\DeclareAbbreviation\Afz{Astrofizika}
\DeclareAbbreviation\AGAb{Astronomische Gesellschaft Abstract Ser.}
\DeclareAbbreviation\an{Astron. Nachr.}
\DeclareAbbreviation\AnAp{Annales d'Astrophysique}
\DeclareAbbreviation\AnTok{Tokyo Astron. Obs. Annals, Sec. Ser.}
\DeclareAbbreviation\Ap{Astrophysics}
\DeclareAbbreviation\ARep{Astron. Rep.}
\DeclareAbbreviation\AstBu{Astrophys. Bull.}
\DeclareAbbreviation\ATel{Astron. Telegram}
\DeclareAbbreviation\ATsir{Astron. Tsirk.}
\DeclareAbbreviation\AcApS{Acta Astrophys. Sinica}
\DeclareAbbreviation\AstL{Astron. Lett.}
\DeclareAbbreviation\BaltA{Baltic Astron.}
\DeclareAbbreviation\BANS{Bull. of the Astron. Institutes of the Netherlands Suppl. Ser.}
\DeclareAbbreviation\BASI{Bull. Astron. Soc. India}
\DeclareAbbreviation\BeSN{Be Newslett.}
\DeclareAbbreviation\BHarO{Harvard Coll. Obs. Bull.}
\DeclareAbbreviation\CBET{Cent. Bur. Electron. Telegrams}
\DeclareAbbreviation\ChJAA{Chinese J. of Astron. and Astrophys.}
\DeclareAbbreviation\caa{Chinese J. of Astron. and Astrophys.}
\DeclareAbbreviation\CoAsi{Asiago Contr.}
\DeclareAbbreviation\CoSka{Contributions of the Astronomical Observatory Skalnat\'e Pleso}
\DeclareAbbreviation\GCN{GRB Coord. Netw. Circ.}
\DeclareAbbreviation\ErgAN{Erg. Astron. Nachr.}
\DeclareAbbreviation\ibvs{IBVS}
\DeclareAbbreviation\IEEEP{IEEE Proc.}
\DeclareAbbreviation\JAD{J. Astron. Data}
\DeclareAbbreviation\JApA{J. of Astrophys. and Astron.}
\DeclareAbbreviation\JAVSO{J. American Assoc. Variable Star Obs.}
\DeclareAbbreviation\JBAA{J. Br. Astron. Assoc.}
\DeclareAbbreviation\JPhCS{J. of Physics Conference Series}
\DeclareAbbreviation\JPSJ{J. Phys. Soc. Japan}
\DeclareAbbreviation\JSARA{J. of the Southeastern Assoc. for Research in Astron.}
\DeclareAbbreviation\LowOB{Lowell Obs. Bull.}
\DeclareAbbreviation\MitAG{Mitteil. der Astronom. Gesell. Hamburg}
\DeclareAbbreviation\MitVS{Mitteil. Ver\"{a}nderl. Sterne}
\DeclareAbbreviation\MmSAI{Mem. Soc. Astron. Ital.}
\DeclareAbbreviation\memsai{Mem. Soc. Astron. Ital.}
\DeclareAbbreviation\Msngr{Messenger}
\DeclareAbbreviation\NewA{New Astron.}
\DeclareAbbreviation\na{New Astron.}
\DeclareAbbreviation\NewAR{New Astron. Rev.}
\DeclareAbbreviation\nar{New Astron. Rev.}
\DeclareAbbreviation\NInfo{Nauchnye Informatsii}
\DeclareAbbreviation\NPhS{Nature Physical Science}
\DeclareAbbreviation\OAP{Odessa Astron. Publ.}
\DeclareAbbreviation\Obs{Observatory}
\DeclareAbbreviation\OEJV{Open Eur. J. on Variable Stars}
\DeclareAbbreviation\PASA{Publ. Astron. Soc. Australia}
\DeclareAbbreviation\PASAu{Publ. Astron. Soc. Australia}
\DeclareAbbreviation\PAZh{Pis'ma AZh}
\DeclareAbbreviation\PJAB{Proc. Japan Acad. Ser. B}
\DeclareAbbreviation\POBeo{Publ. de l'Observatoire Astronomique de Beograd}
\DeclareAbbreviation\PCCP{Phys. Chem. Chem. Phys.}
\DeclareAbbreviation\PhR{Phys. Rep.}
\DeclareAbbreviation\PVSS{Publ. Variable Stars Sect. R. Astron. Soc. New Zealand}
\DeclareAbbreviation\PZ{Perem. Zvezdy}
\DeclareAbbreviation\PZP{Perem. Zvezdy, Prilozh.}
\DeclareAbbreviation\QJRAS{QJRAS}
\DeclareAbbreviation\RA{Ricerche Astronomiche}
\DeclareAbbreviation\RMxAA{Rev. Mexicana Astron. Astrof.}
\DeclareAbbreviation\RvMA{Reviews of Modern Astron.}
\DeclareAbbreviation\SASS{Society for Astronom. Sciences Ann. Symp.}
\DeclareAbbreviation\Sci{Science}
\DeclareAbbreviation\SPIE{SPIE Proc.}
\DeclareAbbreviation\SvA{Soviet Astronomy}
\DeclareAbbreviation\SvAL{Soviet Astronomy Letters}
\DeclareAbbreviation\VeSon{Ver\"{o}ff. Sternw. Sonneberg}
\DeclareAbbreviation\VSOLJBul{VSOLJ Variable Star Bull.}
\DeclareAbbreviation\yCat{VizieR Online Data Catalog}
\DeclareAbbreviation\ZA{Z. Astrophys.}

\def\PublisherCambridge{Cambridge: Cambridge University Press}

\def\PublisherUAP{Tokyo: Universal Academy Press}

\begin{document}
\SetRunningHead{T. Kato and H. Maehara}{Analysis of KIC 8751494}

\Received{201X/XX/XX}
\Accepted{201X/XX/XX}

\title{Analysis of a Kepler Light Curve of the Novalike Cataclysmic Variable
       KIC 8751494}

\author{Taichi \textsc{Kato}}
\affil{Department of Astronomy, Kyoto University,
       Sakyo-ku, Kyoto 606-8502}
\email{tkato@kusastro.kyoto-u.ac.jp}

\and

\author{Hiroyuki \textsc{Maehara}}
\affil{Kiso Observatory,
Institute of Astronomy, School of Science, The University of Tokyo
10762-30, Mitake, Kiso-machi, Kiso-gun,
Nagano 397-0101}
\email{maehara@kiso.ioa.s.u-tokyo.ac.jp}


\KeyWords{accretion, accretion disks
          --- stars: dwarf novae
          --- stars: individual (KIC 8751494)
          --- stars: novae, cataclysmic variables
         }

\maketitle

\begin{abstract}
   We analyzed a Kepler light curve of KIC 8751494,
a recently recognized novalike cataclysmic variable in
the Kepler field.  We detected a stable periodicity of
0.114379(1)~d, which we identified as being the binary's orbital period.
The stronger photometric period around 0.12245 d, which had been
detected from the ground-based observation, was found to be
variable, and we identified this period as being the positive
superhump period.  This superhump period showed short-term 
(10--20~d) and strong variations in period most unexpectedly
when the object entered a slightly faint state.  The fractional
superhump excess varied as large as $\sim$30\%.
The variation of the period very well traced the variation of 
the brightness of the system.  The time-scales of this variation of 
the superhump period was too slow to be interpreted as
the variation caused by the change in the disk radius due to
the thermal disk instability.  We interpreted that the period variation
was caused by the varying pressure effect on the period of
positive superhumps.  This finding suggests that the pressure
effect, in at least novalike systems, plays a very important
(up to $\sim$30\% in the precession rate) role in producing 
the period of the positive superhumps.
We also described a possible detection of the negative superhumps
with a varying period of 0.1071--0.1081~d in the Q14 run of
the Kepler data, and found that the variation of the frequency of the
negative superhumps followed that of positive superhumps.
The relation between the fractional superhump excesses
of negative and positive superhumps can be understood
if the angular frequency of the positive superhumps is decreased 
by a pressure effect.
We also found that the phase of the velocity variation of
the emission lines reported in the earlier study is compatible with
the SW Sex-type classification.
Further, we introduced a new two-dimentional period analysis using
least absolute shrinkage and selection operator (Lasso)
and showed superior advantage of this method.
\end{abstract}

\section{Introduction}

   Cataclysmic variables (CVs) are close binary systems consisting
of a white dwarf and a mass-transferring red dwarf star.
The accreted matter forms an accretion disk around the white
dwarf unless the magnetic field of the white dwarf is sufficiently
strong.  CVs with accretion disks are classified into dwarf novae (DNe),
which show outbursts with amplitudes of 2--8 mag, and novalikes, 
which do not show strong outbursts.  The distinction between
these systems are believed to be a result of the thermal instability
in the accretion disk, and novalike variables are regarded
as systems having higher mass-transfer rates than in DNe,
resulting in thermally stable accretion disks
(see e.g. \cite{osa96review}).

   Superhumps are variations whose periods are 
longer (these superhumps are therefore also called positive superhumps)
than the orbital period by a few percent, and are observed during
superoutbursts of SU UMa-type dwarf novae (a class of dwarf novae)
and in some novalike variables (\cite{pat91v603aql}; \cite{pat99SH}).
These superhumps are believed to arise from the precessing
non-axisymmetric disk whose eccentricity is produced by
the tidal instability arising from the 3:1 resonance
(\cite{whi88tidal}; \cite{hir90SHexcess}).  The fractional
superhump excess ($\epsilon \equiv P_{\rm SH}/P_{\rm orb}-1$ in period
and $\epsilon^* \equiv 1-P_{\rm orb}/P_{\rm SH}$ in frequency,
where $P_{\rm SH}$ and $P_{\rm orb}$ are the superhump period
and the orbital period, respectively) is an observational measure
of the precession rate of the accretion disk.
If we treat the precessing disk dynamically, and ignore
hydrodynamical effects, $\epsilon^*$ is shown to has a form 
\citep{osa85SHexcess}:
\begin{equation}
\epsilon^*=\frac{\omega_{\rm prec}}{\omega_{\rm orb}}
= \frac{3}{4} \frac{q}{\sqrt{1+q}} \left( \frac{R_d}{A} \right)^{3/2},
\label{equ:apprecession}
\end{equation}
where $q = M_2/M_1$ is the mass ratio of the binary, $\omega_{\rm prec}$
$\omega_{\rm orb}$ are the precessional and orbital angular velocity,
and $R_d$ and $A$ are the disk radius and the binary
separation, respectively.
This value was reported to larger than the observed
values (e.g. \cite{mol92SHexcess}) and
it has been shown that the pressure also affect
$\omega_{\rm prec}$ [see \citet{lub92SH}; \citet{hir93SHperiod}.
\citet{mur98SH}; \citet{mon01SH}; \citet{pea06SH} also discuss
implications].  The pressure effect, however, has been usually
regarded as of secondary importance for analysis superhump data
in dwarf novae because the order of a magnitude estimate suggested
that the gravitational effect of the secondary is much
larger than the pressure effect (see \cite{osa85SHexcess}).
We here report the discovery of rapid period variation
of the period of positive superhumps in a novalike variable
in the Kepler (\cite{bor10Keplerfirst}; \cite{Kepler}) field, 
whose variation can be better understood
as a result of the pressure effect.

   This object [KIC 8751494; since the object is located at
\timeform{19h24m10.81s}, \timeform{+44D59'34.9''}, it
is also referred to as KIC J192410.81$+$445934.9 \citep{wil10j1924};
we hereafter use an abbreviation KIC J1924] is
a novalike CV discovered in the Kepler field \citep{wil10j1924}.
Using ground-based photometry and spectroscopy, \citet{wil10j1924}
identified an orbital period of 0.12233(7)~d.  Based on the
narrow hydrogen emission lines and the presence of strong
He\textsc{II} and C\textsc{III}/N\textsc{III} emission lines,
\citet{wil10j1924} suggested that this object may be a member
of SW Sex stars [for a recent summary of SW Sex stars, see
e.g. \citet{rod07swsex}, \citet{rod07newswsex}].

   Since the Kepler data greatly improved our understanding in
KIC J1924, we first give the characterization of this system in
section \ref{sec:basicproperties} and then describe our
interpretation of the period variation of the positive superhumps
in section \ref{sec:shvar}.

\section{Characterization of KIC J1924}\label{sec:basicproperties}

\subsection{Long-Term Variation}

\begin{figure*}
  \begin{center}
    \FigureFile(160mm,80mm){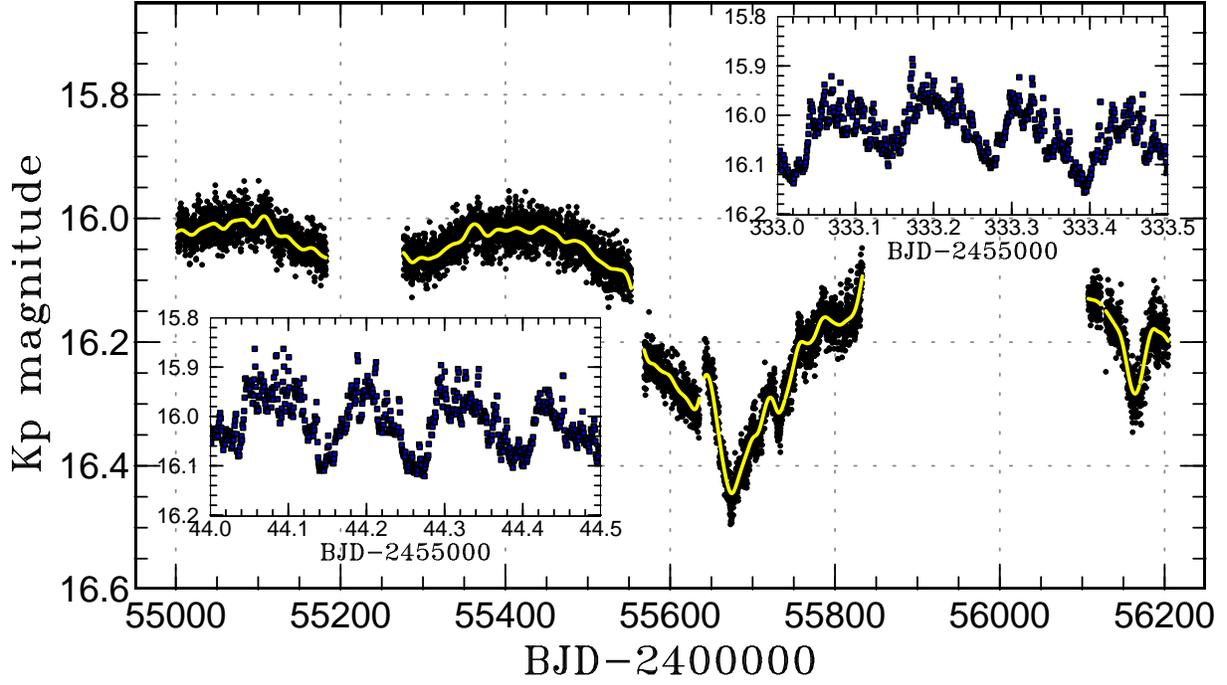}
  \end{center}
  \caption{Long-term light curve of KIC J1924.  Both SC and LC
  data were used.  The data were averaged to bins having widths
  of the orbital period.  The Kepler (Kp) magnitudes were estimated
  while assuming magnitude 12 for 10$^7$ electrons for an SC
  \citep{can10v344lyr}.  There are no globally stable zero-points
  in the Kepler data, and the light curves show discontinuous jumps
  or drops between different quarters.
  In the middle of the points the long-term trend is overplotted.
  Two representative enlarged SC light curves are shown in the insets.
  The modulations seen in these SC light curves are mainly superhumps.
  }
  \label{fig:j1924lt}
\end{figure*}

   We used Kepler public SC data (parts of Q2 and Q3 quarters and
the full Q5 quarter) and LC data (Q2, Q3, Q5--Q10 and Q14) for analysis.
The long-term light curve (figure \ref{fig:j1924lt})
indicates that the object entered
a slightly low state 0.5~mag (2011 April--May) below the brightest 
observations in Kepler.  The object showed a gradual
decline to this minimum since the later half of 2010 and was
slowly rising from this minimum until the end of the observation
(2011 September).  The fading was very shallow compared to
those of VY Scl-type objects (cf. \cite{gre98vyscl}; \cite{lea99vyscl}).
The SC data were obtained around the brightest epochs in the
available Kepler data.

   This object has a close apparent companion ($g$-magnitude 16.71
and separation \timeform{6.8''} according to the Kepler Input Catalog).
Because the Kepler has a pixel size of
\timeform{4''} and is defocused to obtain better a signal-to-noise
ratio for brighter objects, this close companion contaminates
the observed synthetic aperture counts.  We estimated this effect
by measuring the variation of the centroid coordinates of KIC J1924
toward the direction of the companion.
The measured variation was \timeform{0.4''}.  This value and the
direction of the centroid's movement can be well explained
assuming that the companion contributes 1100 electrons s$^{-1}$
to the centroid's variation.
This estimate of the contribution of the companion indicates
that the real range of variation of KIC J1924 was 0.65 mag
assuming that the companion has a constant magnitude.

\begin{figure}
  \begin{center}
    \FigureFile(80mm,100mm){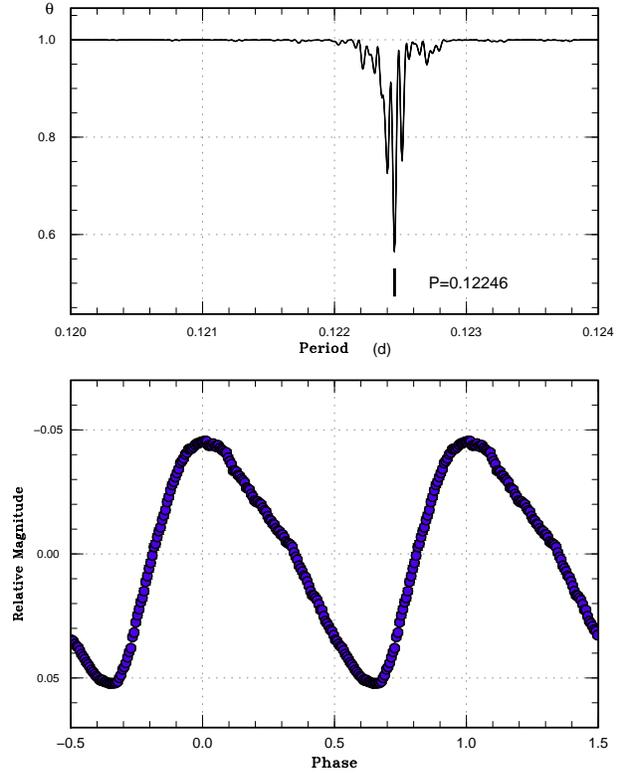}
  \end{center}
  \caption{Mean superhump profile of KIC J1924 from the SC data.
     (Upper): PDM analysis.
     (Lower): Phase-averaged profile.}
  \label{fig:j1924sh}
\end{figure}

\begin{figure}
  \begin{center}
    \FigureFile(80mm,100mm){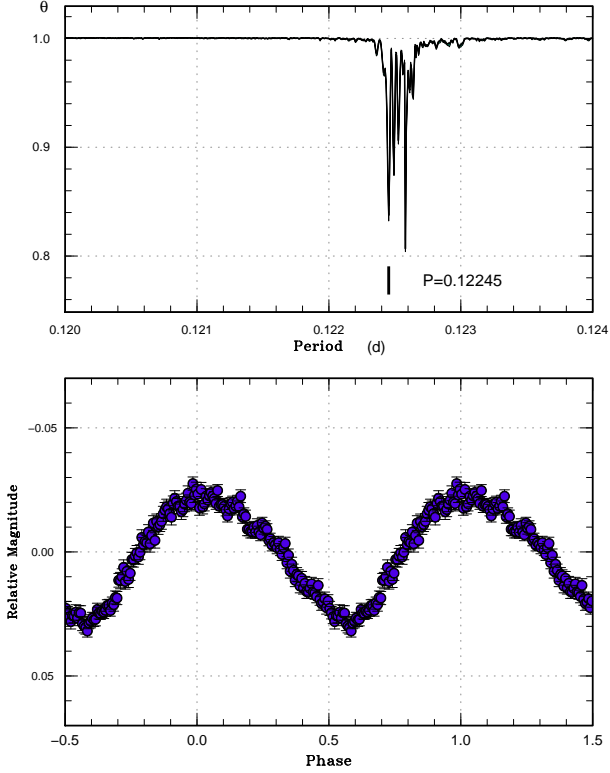}
  \end{center}
  \caption{Mean superhump profile of KIC J1924 from the LC data.
     (Upper): PDM analysis.  The period at 0.12258~d is an artificial
     signal caused by 1/6 of the intervals of LC data.
     (Lower): Phase-averaged profile.}
  \label{fig:j1924shlc}
\end{figure}

\begin{figure}
  \begin{center}
    \FigureFile(80mm,100mm){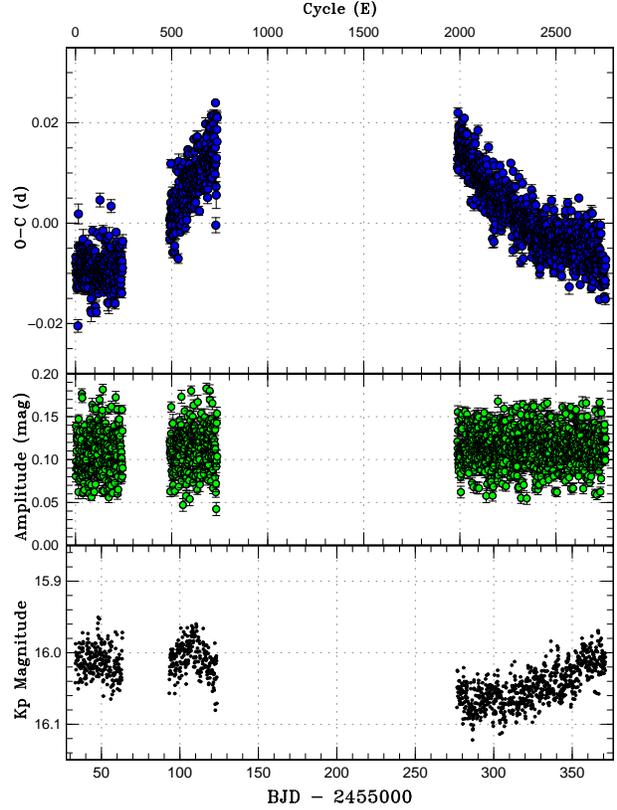}
  \end{center}
  \caption{$O-C$ diagram of superhumps in KIC J1924 in Kepler SC data.
     (Upper): $O-C$ diagram.  The residuals were shown against
     Max(BJD) = $2455033.500+0.122455 E$.  In extracting the times
     of superhump maxima, we used a method in \citet{Pdot} using
     a mean profile of superhumps in KIC J1924.
     (Middle): Amplitudes.  We used $\pm$0.4 phase around the maximum
     for fitting, and the amplitudes shown on the figures are not
     maximum-to-minimum difference.
     (Lower): Light curve.  Kepler magnitudes were averaged to
     each superhump cycle.}
  \label{fig:j1924shoc}
\end{figure}

\begin{figure}
  \begin{center}
    \FigureFile(80mm,100mm){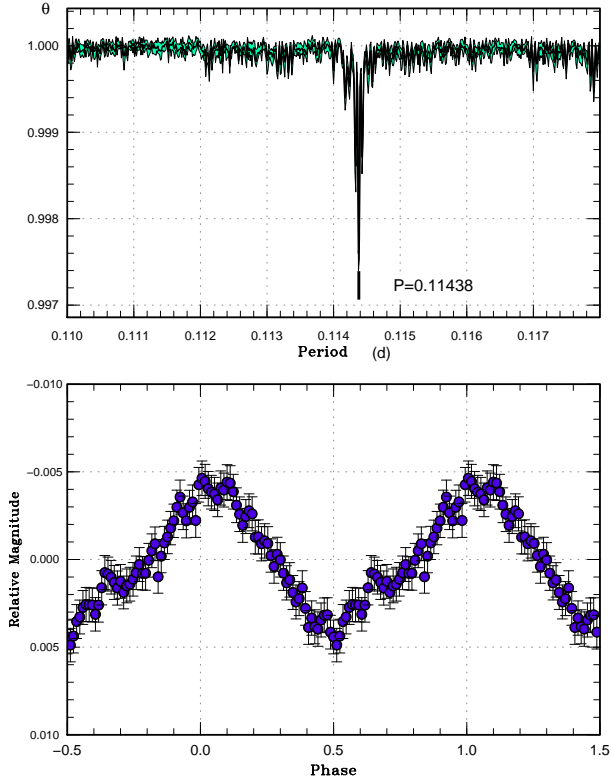}
  \end{center}
  \caption{Orbital variation of KIC J1924 from the SC data.
     (Upper): PDM analysis.
     (Lower): Phase-averaged profile.}
  \label{fig:j1924orb}
\end{figure}

\begin{table*}
\caption{Detected periods in KID J1924 in each quarter.}\label{tab:j1924detper}
\begin{center}
\begin{tabular}{cccccccc}
\hline
cadence type & cadence & \multicolumn{2}{c}{BJD$-$2455000} & $P_1$\commenta & amp\commentb & $P_2$\commenta & amp\commentb \\
             &         & start & end & & \\
\hline
SC & Q2  &  33.32 &  63.30 & 0.1224567(11) & 438 & 0.114279(26) & 58 \\
SC & Q3  &  93.21 & 123.56 & 0.1225151(10) & 424 & 0.114319(18) & 57 \\
SC & Q5  & 276.48 & 371.17 & 0.1224255(2)  & 432 & 0.114376(13) & 35 \\
LC & Q2  &   2.52 &  91.47 & 0.1224689(9)  & 395 & --\commentc & 56 \\
LC & Q3  &  93.22 & 182.50 & 0.1225044(7)  & 426 & --\commentc & 67 \\
LC & Q5  & 276.49 & 371.16 & 0.1224262(7)  & 432 & --\commentc & 80 \\
LC & Q6  & 372.47 & 462.30 & 0.1224551(7)  & 492 & --\commentc & 71 \\
LC & Q7  & 463.17 & 523.23 & 0.1224923(13) & 482 & --\commentc & 124 \\
LC & Q8  & 524.15 & 635.34 & 0.1225825(2)\commenta & 436 & --\commentc & 52 \\ 
LC & Q9  & 641.52 & 738.93 & --\commentd & --\commentd & --\commentc & 38 \\
LC & Q10 & 739.85 & 833.27 & 0.1225383(5) & 229 & --\commentc & 52 \\
LC & Q14 & 1107.14 & 1204.32 & --\commentd & --\commentd & --\commentc & 64 \\
LC & Q2$+$Q3 & 2.52 & 182.50 &          & & 0.114376(9) & \\
LC & Q5$+$Q6 & 276.48 & 462.30 &        & & 0.114368(12) & \\
LC & Q7$+$Q8 & 463.17 & 635.34 &        & & 0.114374(9) & \\
LC & Q9$+$Q10 & 641.52 & 833.27 &       & & 0.114379(8) & \\
SC & all & 33.32 & 371.17 & 0.1224555(1) & & 0.1143794(7) & \\
LC & all & 2.52 & 1204.32 & 0.1224522(4) & & 0.114378(3) & \\
\hline
\multicolumn{8}{c}{\parbox{480pt}{\commenta The values in the
parentheses are 1 $\sigma$ errors.}} \\
\multicolumn{8}{c}{\parbox{480pt}{\commentb Amplitude in electrons s$^{-1}$.
In determining the amplitudes of $P_2$ in the LC data, we assumed a period
of 0.114379~d.}} \\
\multicolumn{8}{c}{\parbox{480pt}{\commentc Although the signal was 
present, the period was not well determined due to the interference 
by the Kepler sampling frequency.}} \\
\multicolumn{8}{c}{\parbox{480pt}{\commentd The period or amplitude
could not be measured due to the strong variation of the period during
this quarter.}} \\
\\
\end{tabular}
\end{center}
\end{table*}

\subsection{Orbital Period and Superhump Period}

   After removing long-term trends, a phase dispersion minimization
(PDM; \cite{PDM}) analysis yielded
the strongest signal at a period (hereafter $P_1$) of 
0.1224555(1)~d (SC data, figure \ref{fig:j1924sh})
and 0.1224533(1)~d (LC data, figure \ref{fig:j1924shlc}).
Although this signal is similar to the one recorded by
\citet{wil10j1924}, we did not regard this signal as
the orbital period, because there are a variation in period
between quarters (table \ref{tab:j1924detper}) and a large
systematic $O-C$ variation against a constant period
(figure \ref{fig:j1924shoc}).  We regard the strongest signal
as the superhump period ($P_{\rm SH}$), 
whose possibility was already mentioned 
in \citet{wil10j1924}.

   There is the second strongest signal the period of which is
$\sim$7\% shorter than $P_1$.  We call this period $P_2$.  
This period was detected in individual SC runs.  Although the period was
present in the individual LC runs, the period was difficult to
measure because of interference affected by nearly constant 
sampling intervals in Kepler.  When two adjacent LC runs were
combined, we could also measure $P_2$ in LC runs
(table \ref{tab:j1924detper}; the periods were determined
with the PDM method and the errors are 1 $\sigma$ error by
\cite{fer89error}, \cite{Pdot2}).  Since $P_1$ could be measured in
most individual LC runs, we did not measure $P_1$ for the combined
adjacent LC runs.  For comparison, we also listed $P_1$ and $P_2$
measured using all of the data.  These measurements indicate
that $P_2$ was constant between different quarters and we regard
this period as the orbital period.
The period is in good agreement with the ones, derived from two
combined LC and SC data
(LC data have an advantage of larger cycle numbers, while
SC data have higher time resolution, and they are complementary),
we adopted a period of 0.114379(1)~d by combining these measurements.
The mean orbital profile based on the SC data is shown in
figure \ref{fig:j1924orb}.  The profile is more symmetric
than that of the superhumps.  The epoch of the maximum brightness
corresponds to BJD 2455228.2463 (the epoch was chosen as the nearest
one to the average of the SC observation).

   The mean amplitudes of $P_1$ and $P_2$ during Q2--Q7 (bright state)
were 0.10~mag (430 electrons s$^{-1}$) and 0.014 mag
(57 electrons s$^{-1}$).  The mean amplitudes of these signals
in individual quarters are given in table \ref{tab:j1924detper}.
The resultant periods yielded a fractional superhump excess
$\epsilon$ of 7.06\% (mean value).

\begin{figure}
  \begin{center}
    \FigureFile(88mm,110mm){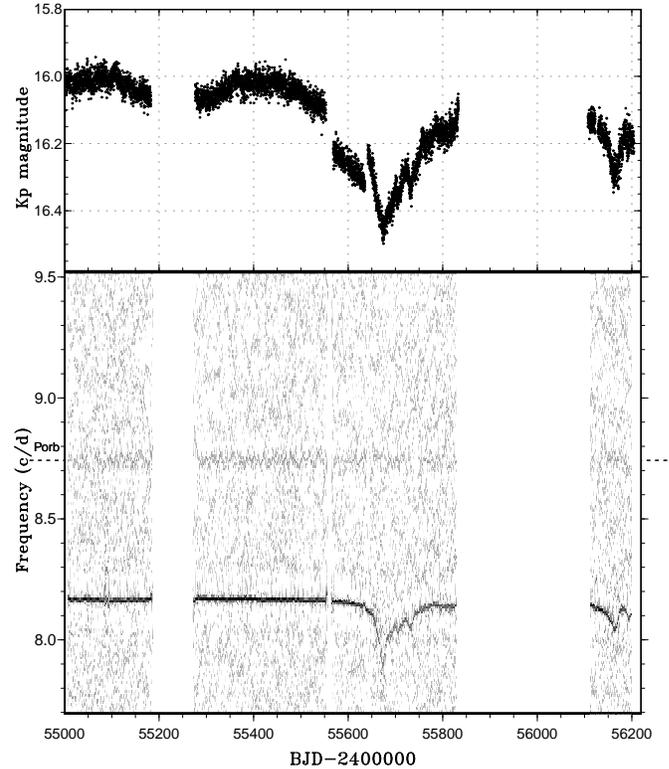}
  \end{center}
  \caption{Lasso 2-D power spectrum analysis of KIC J1924 from the LC data.
     (Upper): Light curve.  The Kepler data were binned to
     the mean superhump period.
     (Lower): Result of the lasso analysis ($\log \lambda=-4.8$).
     The superhump period was discovered to be a signal with a variable
     frequency, while the orbital period was discovered to be a constant
     frequency.}
  \label{fig:j1924lassopowlc}
\end{figure}

\subsection{Two-Dimentional Period Analysis using
Least Absolute Shrinkage and Selection Operator (Lasso)}

   We here introduce a new method to calculate two-dimensional 
power spectra using least absolute shrinkage and selection operator
(Lasso, \cite{lasso}), which was introduced to analysis of astronomical 
time-series data \citep{kat12perlasso}.
In Fourier-type period analysis, there is the well-known Heisenberg-Gabor
limit, that is, one cannot simultaneously localize a signal 
in both the time domain and the frequency domain; this characteristic
of Fourier-type period analysis is disadvantageous to the analysis
of such rapidly changing periods, as superhumps in outbursting 
dwarf novae have.
The advantage of the Lasso analysis is that this method is not
restricted by the Heisenberg-Gabor limit and the peaks 
in the power spectra are very sharp.\footnote{
   A simple explanation why the Lasso-type analysis can overcome
   the Heisenberg-Gabor limit in a different field of science
   can be found in \citet{her09radarCompressedSensing}.
}
There is, however, a set-back of this method that 
the resultant power is not linear in power amplitude. 
It is very advantageous to use the Lasso period analysis 
in resolving closely spaced signals when we know that only 
a small number of signals coexist.  Lasso gives very sharp signals 
in the two-dimensional power spectra, and thus it is suitable 
for studying the signals with rapidly varying frequency, 
such as superhumps in outbursting 
SU UMa-type dwarf novae (e.g. \cite{osa13v344lyrv1504cyg}).

   The power spectra were estimated
obtained by applying the method of \citet{kat12perlasso} to 
10~d bins, which were shifted in a step of 1~d
after removing the long-term trend using locally-weighted polynomial 
regression (LOWESS, \cite{LOWESS}).
The 200 frequencies were evenly spaced
between frequencies 7.69 and 9.52 c/d to extract the spectra.

\begin{figure*}
  \begin{center}
    \FigureFile(88mm,110mm){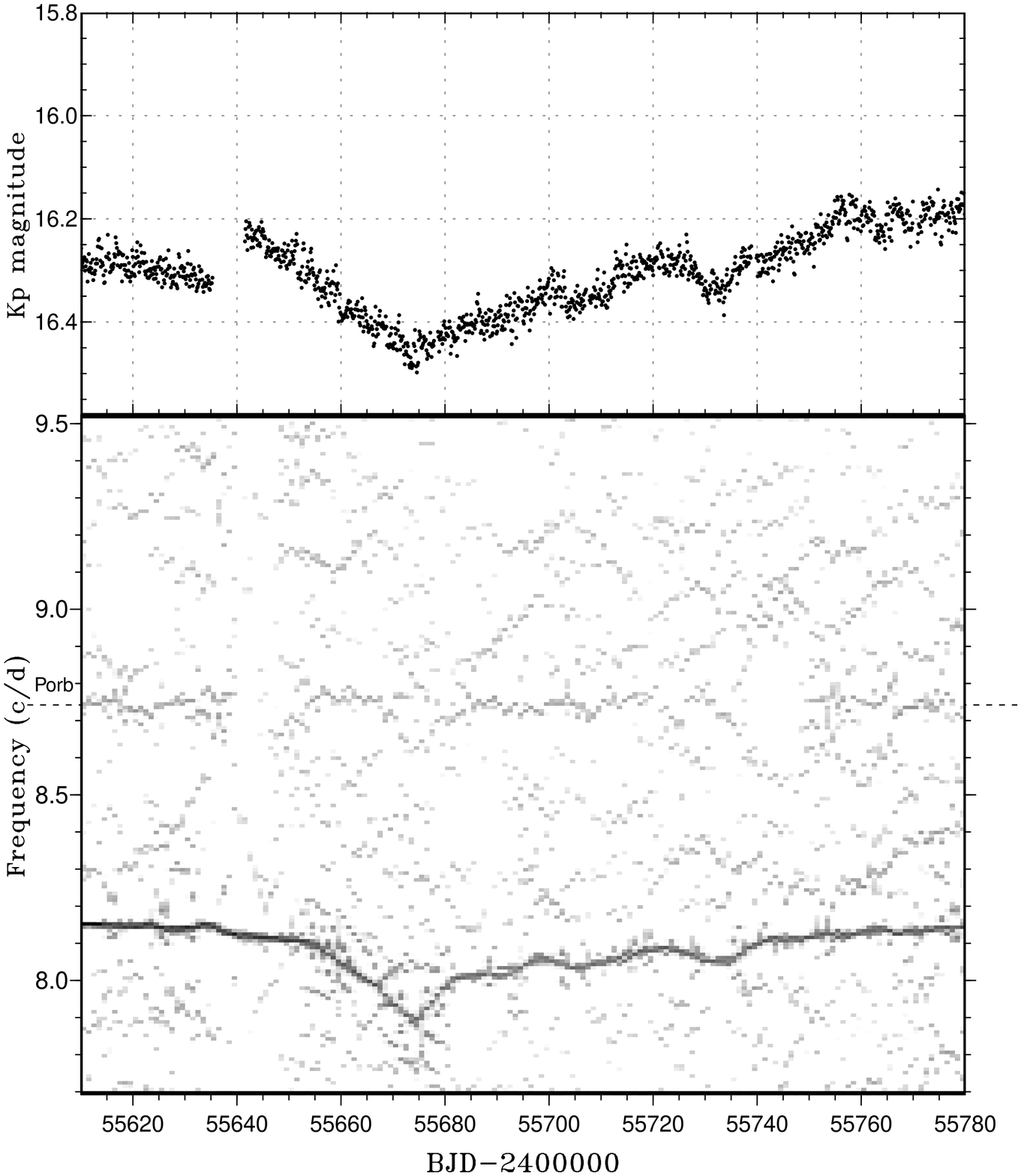}
    \FigureFile(88mm,110mm){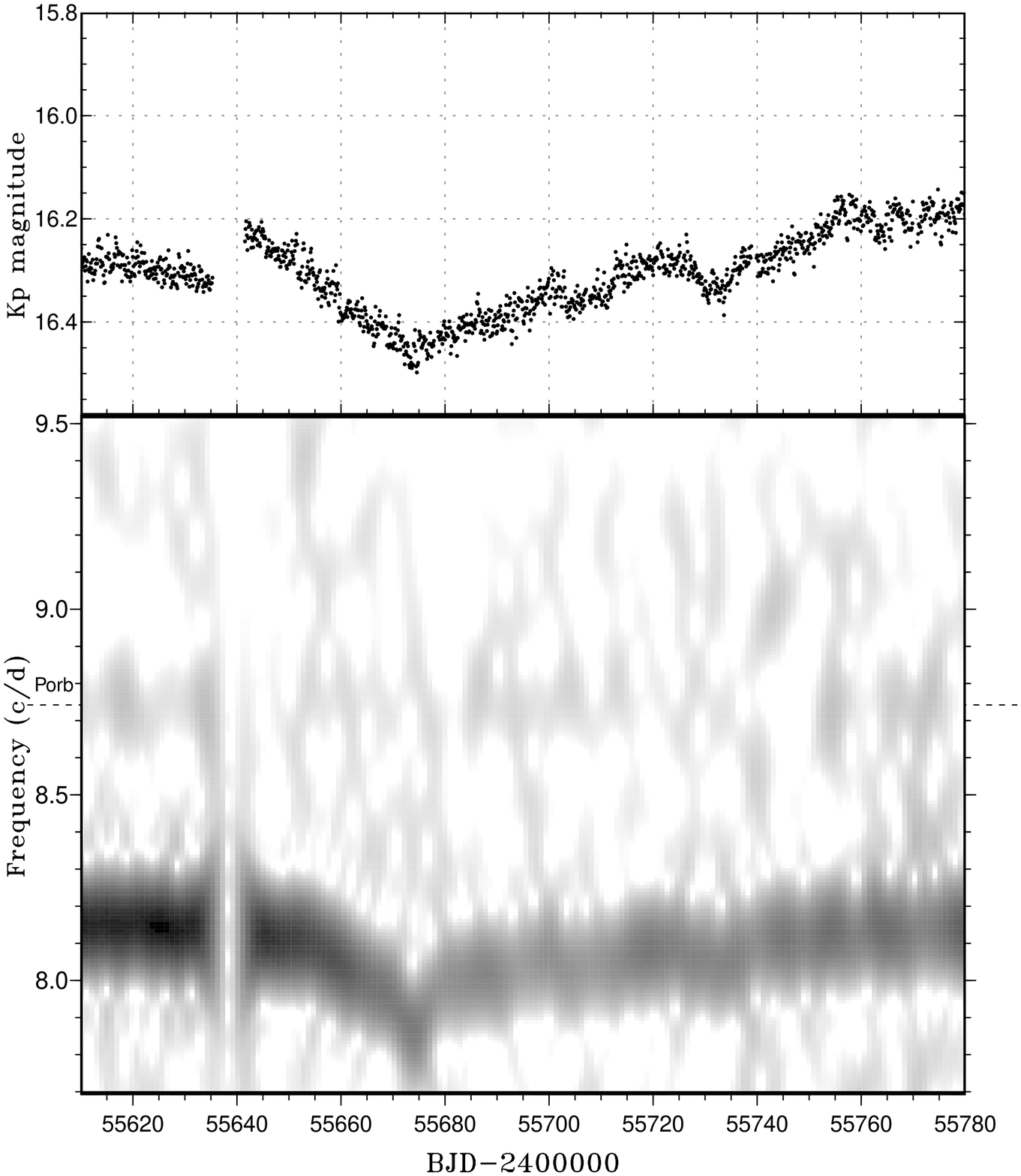}
  \end{center}
  \caption{(Left) Enlargement of figure \ref{fig:j1924lassopowlc}.
     Short-term variations of the superhump period during the
     fading are clearly seen.
     (Right) Discrete Fourier transform of the same data.
     The advantage of Lasso over Fourier is clearly seen.}
  \label{fig:j1924lassopowlc2}
\end{figure*}

   There is a free parameter $\lambda$ in giving an $\ell_1$ term
(cf. \cite{kat12perlasso}).  In producing figures
\ref{fig:j1924lassopowlc}, \ref{fig:j1924lassopowlc2},
we selected $\lambda$ which gives the signals
best contrasted against a background, i.e.
the most physically meaningful parameter.  This paramter $\lambda$ 
is close to the most regularized model with a cross-validation error
within one or two standard deviations of the minimum
(cf. R Sct for \cite{kat12perlasso}).
We also applied smearing of the signals between $\pm$3 bins
shifted in a step of 1~d while considering the width of the window. 

   The resultant spectrum is shown in figures 
\ref{fig:j1924lassopowlc} and \ref{fig:j1924lassopowlc2}.
The latter figure also shows a comparison between Lasso and
Fourier analyses.
A similar analysis to the SC data yielded a similar result
with a better time-resolution.  Since there was no additional
remarkable feature in the analysis of the SC data, 
we omitted this figure.

\subsection{KIC J1924 as an SW Sex Star}

   Although it is not clear which orbital phase corresponds
to the optical maximum of the orbital variation, it can be
the phase of the the superior conjunction of the secondary 
if the maximum represents the reflection effect.  We therefore 
define here the orbital phase of the optical maximuma as 0.5 to make
a comparison with eclipsing SW Sex stars.
The spectroscopic zero phase (red-to-blue crossing
of the velocities of the emission lines)
$T_{0,{\rm spec}}$ = 24454734.66646(383)
\citep{wil10j1924} corresponds to an orbital phase of
0.20 (the accumulative error of the phase resulting from 
the error in the period is smaller than 0.04 even assuming
an extreme error of 10$^{-5}$~d).
If the emission line traces the motion of the white dwarf,
this phase should be close to zero, and this phase delay
of the Balmer lines appears to qualify one of the SW Sex-type
properties.

   The optical spectrum of KIC J1924 is characterized by broad
absorption lines apparently arising from the optically thick
accretion disk (figure 4 in \cite{wil10j1924}).  Although
this kind of ``UX UMa-type'' spectrum [cf. \citet{war95book}
for the type definition; \citet{dhi13swsexenigma} suggests
that whether the lines are seen in emission or in absorption
may be an inclination effect and proposed to combine UX UMa-type
and RW Tri-type with pure emission lines into the UX UMa-type
novalike star] is relatively rarely seen in SW Sex stars
(see figure 2 in \cite{rod07swsex}), there are some known
SW Sex-type object showing the broad Balmer absorption 
component: HL Aqr \citep{hun85hlaqr}, LN UMa (\cite{hil98pg1000};
\cite{rod07newswsex}. A flux calibrated low-resolution
spectrum of KIC J1924 is also shown in \citet{ost10Keplercompactpulsator},
which can be more directly comparable to other SW Sex-type stars.

   Improved time-resolved optical spectroscopy, which will 
provide an independent measure of the orbital period would test 
the results of the paper, and provide a more solid classification
of this system.

\subsection{KIC J1924 as a Permanent Superhumper}

   KIC J1924 persistently showed positive superhumps, during at least
Kepler observations up to 2011 September.
The object is thus classified as a permanent superhumper.
Since the same superhumps were observed in \citet{wil10j1924},
the positive superhumps seem to have been very stably seen
in this object.  The object appears to be very similar to V795 Her,
another novalike system in the period gap having a similar
orbital period (\cite{pat94v795her}; \cite{pap06v1193orilqpegv795her};
\cite{sim12v795her}).  The superhumps of KIC J1924, however, may be
more stable than those in V795 Her (\cite{pat94v795her}; 
\cite{pap06v1193orilqpegv795her}).

   In table \ref{tab:persuperhumper}, we list permanent superhumpers
having orbital periods similar to that of KIC J1942.  The recorded
$\epsilon$=7.1\% is the characteristic value of this $P_{\rm orb}$.
Among these objects, only V442 Oph is known to show $\sim$3 mag
fading\footnote{
  See $<$http://www.astrosurf.com/blazar/variable/UG04/\\V442\%20Oph.html$>$.
}
comparble to typical VY Scl-type variables
[see light curves in e.g. \citet{gre98vyscl}, \citet{lea99vyscl}].

\begin{table*}
\caption{Permanent superhumpers with orbital periods similar to
that of KIC J1942.}\label{tab:persuperhumper}
\begin{center}
\begin{tabular}{ccccccc}
\hline
Object    & $P_{\rm orb}$ & $P_{\rm SH}$ (positive) & $P_{\rm SH}$ (negative) & 
$\epsilon$ (\%) & SW Sex-type & References \\
\hline
AH Men    & 0.12721  & 0.1385   &   --    & 8.9 & Yes & 1,2 \\
V442 Oph  & 0.12433  &    --    & 0.12090 & $-$2.8 & Yes & 3,4,5 \\
V1084 Her & 0.12056  &    --    & 0.11696 & $-$3.0 & Yes & 5,6 \\
V592 Cas  & 0.115063 & 0.12228  &   --    & 6.3 & No & 7 \\
KIC J1942 & 0.114379 & 0.12245\commenta & 0.1071--0.1081 & 7.1/$-5.5\sim-6.4$ & Yes & this work \\
V795 Her  & 0.108247 & 0.116486 &   --    & 7.6 & Yes & 8,9,10,11 \\
V348 Pup  & 0.101839 & 0.108567 &   --    & 6.6 & Yes & 12,13,14 \\
\hline
\multicolumn{7}{l}{\commenta Variable period.  Mean value.} \\
\multicolumn{7}{c}{\parbox{440pt}{
1: \citet{rod07swsex},
2: \citet{pat95ahmen},
3: \citet{hoa00v442oph},
4: \citet{hoa00v442ophletter},
5: \citet{pat02v442ophj1643},
6: \citet{rod07newswsex},
7: \citet{tay98v592cas},
8: \citet{sha90v795her},
9: \citet{zha91v795her},
10: \citet{cas96v795her},
11: \citet{sim12v795her},
12: \citet{rol00v348pup},
13: \citet{rod01v348pup},
14: \citet{dai10uzforv348pup}
}}
\end{tabular}
\end{center}
\end{table*}

\section{Variation of the Superhump Period}\label{sec:shvar}

\begin{figure}
  \begin{center}
    \FigureFile(88mm,110mm){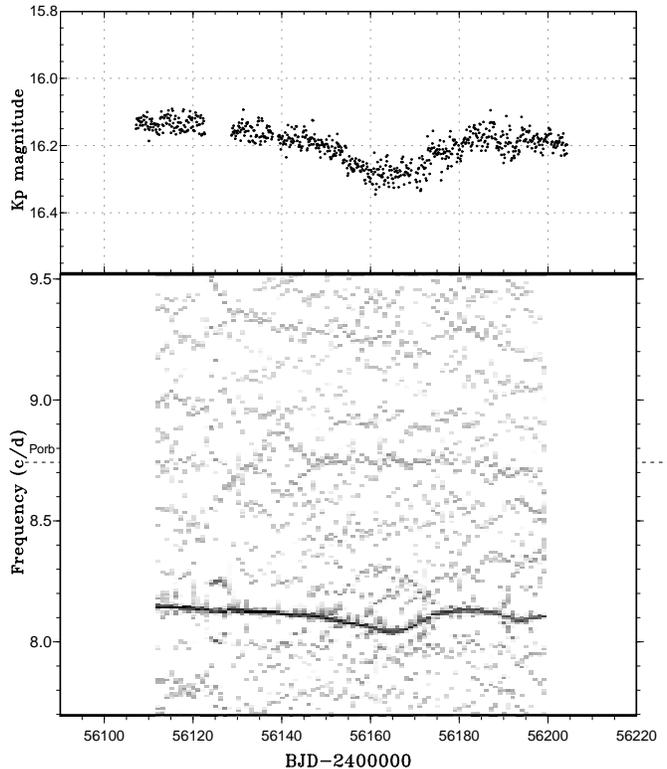}
  \end{center}
  \caption{Period analysis of Q14 data.
     Short-term variations of the superhump period during another
     epoch of fading are clearly seen.
     The signals around frequencies 9.2--9.3 c/d are negative
     superhumps, whose frequencies also varied in accordance to
     the system brightness.}
  \label{fig:j1924lassopowlc3}
\end{figure}

\subsection{Rapid Period Variation -- Radius Variation?}\label{sec:pervar}

   As shown in figure \ref{fig:j1924lassopowlc}, the frequency 
of the superhump was almost constant before BJD 2455200.
After a gap in the Kepler observation, the frequency started to
decrease very gradually.  The frequency, however, suddenly showed
a drop at around BJD 2455620, when the brightness of the system
dropped significantly.  The frequency of the superhump signal
even reached 7.9 c/d (=0.127~d in period), which marked a large
$\epsilon$ of 11\% ($\epsilon^*$ amounted to 10\%).
The obtained frequencies are perfectly
in agreement with those obtained using the PDM, as performed in
\citet{osa13v1504cygKepler}.

   Assuming that $\epsilon^*$ for positive superhumps depends on
$R_{\rm d}^{3/2}$ \citep{osa85SHexcess}, 
this translates to a $\sim$35\% increase in disk radius.
Adopting a mean $\epsilon$ of 7.1\%, we can obtain
$q=0.30$ using the $\epsilon-q$ relation in \citet{Pdot}
[the relation in \citet{pat05SH} yields $q=0.27$].
At this $q$, the tidal truncation radius is close to that
of the 3:1 resonance and the disk is not expected to expand
beyond the tidal truncation radius for a long time.
The radius of the Roche lobe is $\sim 0.50A$.
Considering that the radius of the 3:1 resonance is $\sim 0.46A$,
it is difficult to ascribe period increase solely to
the disk expansion.

\subsection{Pressure Effect}

   We then consider if the dwarf nova-type disk instability
can account for the observed variation of the disk radius.
The rise time of dwarf nova-type outburst can be expressed as
\begin{equation}
\tau_{\rm r}=0.17 \left( \frac{0.1}{\displaystyle \alpha_{\rm H}} \right) M_1^{1/3}(1+q)^{1/3}P_{\rm orb}^{2/3}({\rm h}) \; {\rm d} \quad (P_{\rm orb}\le 9 {\rm h}),
\label{equ:dnrise}
\end{equation}
where $\tau_{\rm r}$ and $\alpha_{\rm H}$ are the rise time
and the viscosity parameter in the hot state, respectively
(equation 3.21 in \cite{war95book}).  By using typical values
of $M_1$=0.8$M_\odot$, $\alpha_{\rm H}$=0.1 and assuming $q$=0.3,
we obtain $\tau_{\rm r}\sim$0.3~d.
The decay time is also expressed as
\begin{equation}
\tau_{\rm d}=0.75 \left( \frac{0.1}{\displaystyle \alpha_{\rm H}} \right) M_1^{2/3}(1+q)^{1/6}P_{\rm orb}^{1/3}({\rm h}) \; {\rm d} \quad (P_{\rm orb}\le 9 {\rm h}),
\label{equ:dndecay}
\end{equation}
where $\tau_{\rm d}$ is the decay time (equation 3.24 in \cite{war95book}).
The value of $\tau_{\rm d}$ for the same set of parameters
is $\sim$0.9~d.  The both time scales are much shorter than
what were seen in subsection \ref{sec:pervar}.
It is thus unlikely that the period variation is caused by
the dwarf nova-type instability.

   In the analysis, up to now we have only used the dynamical
term (by the tidal force of the secondary) of producing the precession
rate, as approximated in \citet{osa85SHexcess}.  It has been
shown that the precession frequency is a sum of the dynamical
term, the pressure effect (or an effect of a finite thickness
of the disk) producing the retrograde precession and a minor
wave-wave interaction term \citep{lub92SH}.  Since the variation
of the observed superhump period very exactly traced the variation of
the system brightness, we suggest that the pressure effect
is responsible for the variation of the period.
This pressure effect is an increasing function of the sound
velocity at the edge of the disk ($c_0$; \cite{hir93SHperiod}),
which has a dependence of $c_0 \propto \sqrt{T_{\rm C}}$, where
$T_{\rm C}$ is the temperature of the midplane of the disk.
If the disk in a novalike variable can be assumed to be a thermally
stable standard disk, the temperature of the midplane has
a dependence of $T_{\rm C} \propto \dot{M}^{3/10}$, where
$\dot{M}$ is the mass-transfer rate (equation 5.49 in
\cite{AccretionPower3}).  The observed luminosity variation
of 0.65 mag translates to an $\sim$9\% variation of $T_{\rm C}$.
By using table 1 in \citet{hir93SHperiod} (although this is for
a $q=0.15$ case, we believe that the dependence of $q$ on
this estimation is small), a comparison between $c_0=0.1$ (realistic
novalike disk in a hot state) and $c_0=0.085$ yielded 
an expected $\epsilon^*$ variation of $\sim$3\%.  Although this
value is smaller than the observed one, this discrepancy
might be resolved if the actual disk in a novalike variable
in a slightly low state may have a deviation from the steady
standard accretion disk.  The importance of the pressure effect
may be partly attributed to the inferred high mass-transfer rates
in SW Sex stars (\cite{rod07swsex}; \cite{tow09CVWDtemp}).
Since most of the permanent superhumpers are SW Sex stars,
this might result a systematic bias if we calibrate 
the $\epsilon-q$ relation using both dwarf novae and permanent
superhumpers.

   Following this logic that the period variation of
the superhumps can be attributed to the pressure effect,
we assumed that the largest $\epsilon^*$ was recorded when
the disk had the oscillation wave limited to the edge of
the accretion disk (corresponding to $c_0 \sim 0$).
The precession rate can be expressed as the following equation
\citep{hir90SHexcess}:
\begin{equation}
\frac{\omega_{\rm prec}}{\omega_{\rm orb}}=\frac{q}{\sqrt{1+q}}
\Bigl[\frac{1}{2}\frac{1}{\sqrt{r}}\frac{d}{dr}\Bigr(r^2\frac{dB_0}{dr}\Bigr)\Bigr],
\label{equ:presfreq}
\end{equation}
where $B_0$ is
\begin{equation}
B_0(r)=b_{1/2}^{(0)}/2=\frac{2}{\pi}\int_0^{\pi/2}\frac{d\phi}{\sqrt{1-r^2\sin^2 \phi}},
\label{equ:laplace}
\end{equation}
which is the Laplace coefficient of the order 0 in celestial mechanics
\citep{sma53Celestialmechanic}.
Numerically calculating this function assuming a radius of
the 3:1 resonance of $r_{3:1}=3^{(-2/3)}(1+q)^{-1/3}$, we could estimate
$q=0.37$ from the largest measured $\epsilon^*$=10\%.
This $q$ is moderately in agreement with the value obtained from
the conventional $\epsilon-q$ relation, and we do not consider that
our interpretation leads to an obvious contradiction.
If the validity of this method is established, observing
period variation of positive superhumps in permanent superhumpers
will be a tool in estimating mass ratios by using superhumps.

   A similar pattern, but with weaker period variations, was
also recorded during the Q14 (figure \ref{fig:j1924lassopowlc3}).

\subsection{Negative Superhumps}

   During the Q14 run, weak signals of negative superhumps appeared to
be present around frequencies of 9.2--9.3 c/d
(figure \ref{fig:j1924lassopowlc3}).  We identified these signals
as real ones, because close, non-overlapping bins show
similar enhancement of signals.  Since these non-overlapping bins
share no observations in common, we regarded them as a signature
for the independent detection.  The signal, however, was weak
and it was extremely difficult to obtain a convincing, phase-averaged
light curve of this signal due to the variation of the frequency
(just like the difficulty in obtaining a period for positive superhumps
with period variation), as discussed below, and due to the coarse LC sampling.
We describe the properties and implications of the signals assuming 
that the signals are real.  This assumption may be tested by
future Kepler observations, especially if SC runs are available.
These negative superhumps were not detected in other quarters.
The frequency of negative superhump varied as in the same manner
as that of positive superhuumps;
the frequency became smaller when the system slightly faded at around
BJD 2455160--2455170 (frequency 9.25 c/d, period 0.1081~d,
$\epsilon^*=-5.8$\%) compared to the brighter state around 
BJD 2456125--2456140 (frequency 9.34 c/d, period 0.1071~d,
$\epsilon^*=-6.8$\%).
The variation followed a very similar pattern to that of
positive superhumps.
Since the pressure effect is not expected to work for producing
negative superhumps (\cite{osa13v1504cygKepler}; see also
more detailed discussion in \cite{osa13v344lyrv1504cyg}), 
we can attribute this frequency variation
to the change in the radius of the disk.  Since $\epsilon^*$
has a dependence of $R_d^{3/2}$ \citep{lar98XBprecession},
this variation of $\epsilon^*$ corresponds to a variation
of 11\% in the radius of the accretion disk.  The observation
of a lower $\epsilon^*$ in a fainter state is consistent with
the picture that a disk is expected to be smaller when the
mass-transfer rate decreases.

   There is a known relationship between
$\epsilon^*_+$ (positive superhumps) and $\epsilon^*_-$
(negative superhumps); that is $\epsilon^*_+ \simeq 7/4 |\epsilon^*_-|$
[from \citet{lar98XBprecession}; see \cite{osa13v344lyrv1504cyg}
for detailed discussion; this factor has been observationally known
to be close to 2, cf. \citet{pat97v603aql}; \citet{woo09negativeSH}]
and clearly the variation in this object
does not follow this relation.  This result
can also be understood if the $\epsilon^*_+$ is suppressed
by the pressure effect when the system is bright.

\subsection{Variation in Mass-Transfer}

   It appears that the mass-transfer rate from the secondary
in KIC J1924 varied significantly on a time-scale of 10--20~d.
These values are much shorter than in typical VY Scl-type
stars (the typical low states lasts for months to years).
\citet{gar88vyscldqher}, however, showed much shorter historical
fadings in V442 Oph.  This and present finding suggest that
mass-transfer rates in novalike systems significantly vary
on a time-scale as short as 10--20~d.  V795 Her, an object
very similar to KIC J1942, did not experience deep low states
such as are seen in typical VY Scl-type stars [\citet{wen88v795her};
this has also been confirmed by recent VSNET \citep{VSNET},
AAVSO and CRTS \citep{CRTS} observations].  V795 Her also
showed low-amplitude (up to 1 mag) long-term variations
such as were seen in \citet{sim12v795her}.  It would be interesting
to see whether the superhump periods in V795 Her varies
in the same way as in KIC J1924.

\section{Conclusion}

   We studied a public Kepler light curve of the novalike
variable KIC 8751494 (KIC J192410.81$+$445934.9).
Although the photometric period (0.12245~d, Kepler data)
identified by the ground-based observation has also 
been confirmed in the Kepler data, we identified it as
the superhump period based on the long-term instability of 
the period.  We alternatively identified a weaker, but a stable
periodicity of 0.114379(1)~d as being the orbital period.
The inferred fractional superhump excess of 7.06\% (mean value)
is typical of the permanent superhump with an orbital period
similar to that of this object.  Based on the refined orbital ephemeris,
we identified a phase shift in the emission lines in a radial
velocity study reported earlier, which strengthens the identification
of this object as one of SW Sex stars.
The period of superhumps showed a very large (up to $\sim$30\%
in fractional superhump excess) variation when the object
faded in 2011 April--May.  The variation of the period almost
exactly traced the variation of the system brightness.
We examined the origin of the variation of the superhump period
assuming the disk-radius variation and dwarf nova-type
disk instability.  Each possibility is difficult to explain
the observed variation.  We alternatively suggest that the
pressure effect in producing the precession rate of
the non-axisymmetric disk plays a more important role.
This finding suggests that the pressure effect can have
an effect up to $\sim$30\% in the precession rate, in at least
novalike systems. 
We also describe a possible detection of negative superhumps
with a varying period of 0.1071--0.1081~d in the Q14 run,
and found that the variation of the frequency of
negative superhumps followed that of positive superhumps.
The relation between the fractional superhump excesses
between negative and positive superhumps can be understood
if the angular frequency of positive superhumps is decreased 
by the pressure effect.

\medskip

We are grateful to Prof. Yoji Osaki for discussions.
This work was supported by the Grant-in-Aid for the Global COE Program
``The Next Generation of Physics, Spun from Universality and Emergence"
from the Ministry of Education, Culture, Sports, Science and Technology
(MEXT) of Japan.
We thank the Kepler Mission team and the data calibration engineers for
making Kepler data available to the public.

\end{document}